# Label-free optical imaging of ion channel activity on living cells

Qing-Yue Li, Pin-Tian Lyu, Bin Kang*, Hong-Yuan Chen, Jing-Juan Xu*

E-mail: binkang@nju.edu.cn (B. K.), xujj@nju.edu.cn (J.-J. Xu )

State Key Laboratory of Analytical Chemistry for Life Science, Chemistry and Biomedicine Innovation Center, School of Chemistry and Chemical Engineering, Nanjing University, Nanjing 210023, China

**Abstract**

Deciphering ion channel activity and signaling interactions within cells is one of the key tasks of neuroscience. Currently, measuring this electrophysiological activity is done using patch-clamp[1-4] or voltage-sensitive imaging[5-8]. Unfortunately, these techniques are unable to balance between single-channel sensitivity and high-throughput detection. Here we introduce a label-free electrochemical-modulated interferometric scattering microscope (EM-iSCAT) to measure ion channel activity on living cells at both whole-cell and single-channel levels. We visualize the cellular responses dynamics to osmotic stimulation, and record open-close trajectories of single N-methyl-D-aspartate receptors channels with a frame rate of 1.5 kHz. Furthermore, we localize and distinguish different kinds of ion channels ($Na^+$, $K^+$, $Ca^{2+}$) on cell membrane and monitor spatio-temporal heterogeneous responses between different cells in a network. The high-throughput and single-channel sensitive nature of EM-iSCAT microscopy allows monitoring simultaneously the activity of individual channels, their localization, and clustering in the cellular community. Our imaging concept opens the possibility to study any kind of ion channels, and more broadly, cell communication mediated by ion channels.

**Main**

Modern electrophysiology provides vital insights into the electrical activity in various biological systems, from a single ion channel to the complex interactions within cell community[9]. Patch-clamp[1-4] and voltage-sensitive imaging[5-8] have been widely used as two major tools in neuro electrophysiology research. Patch-clamp benefits from extremely high sensitivity and microsecond temporal resolution, allowing real-time observation of the activity of single ion channels. However, its detection is restricted by a monitoring area typically <1 μm$^2$ and the lack of spatial information, and it may cause damage to cells during sealing thus limits the recording time. Instead, although voltage-sensitive imaging can achieve high throughput, its sensitivity is insufficient to track single ion channel activity. Moreover, the fluorescent labeling may affect the structure and function of cell membranes and it suffers from photobleaching and phototoxicity. Other methods, such as the patch clamp arrays[10,11] or electrophysiological chips[12,13], although improving throughput, sacrifice measurement sensitivity. In recent years, a variety of label-free optical electrophysiological methods[14-18] have been developed to overcome these obstacles, which achieves great advances in recording and imaging of cell action potential in real-time. However, thus far, no imaging method has been capable of observing the activity of single ion channels due to the low sensitivity or low frame rate.

Here we reported electrochemical-modulated interferometric scattering microscopy (EM-iSCAT) as a high-throughput and ultrasensitive, label-free imaging platform to visualize and record the cellular electrical activity at both whole-cell and single ion-channel levels. The iSCAT signal comes from the changes in the local charge density as well as the refractive index on cell membrane, induced by the ion diffusion field when an ion channel opens. The signal from ion channels is further amplified and extracted using an electrochemical modulation and signal demodulation. We imaged the real-time cellular responses to osmotic stimulation and tracked the open-close events of single N-methyl-D-aspartate receptors (NMDARs). Moreover, we demonstrated the potential applications of EM-iSCAT for the localization and identification of different ion channels ($Na^+$, $K^+$, $Ca^{2+}$) as well as the channel clusters, from a single cell to cell community.

**Principle of EM-iSCAT**

Interferometric imaging has gained its popularity in chemical and biological research for its high sensitivity and sub-nanometer resolution[19,20]. Our imaging approach is based on the principle of iSCAT[21-23]. Using elastic scattering of visible light to achieve fast acquisition time and ultra-sensitivity, iSCAT has been explored for imaging single molecule interaction on cells[24,25], as well as carrier and ion transport in materials[26-28]. Here we combined an electrochemical-modulated technique with iSCAT, named as EM-iSCAT. Briefly, the wide-field iSCAT was operated in a common-path configuration equipped with a partial reflector to enhance the image contrast (Fig. 1a). The cells were cultured on an indium tin oxide (ITO) substrate and an alternating current (AC)-modulated voltage was applied to the entire sample chamber by a three-electrode system. The aim of electrochemical modulation is to selectively extract the charge density signals and also improve the signal-to-noise ratio. Note that this modulated voltage did not change the membrane potential or alter the ion channel activity (see SI section 4 for details). When ion channels on the cell membrane open, typically $10^6$-$10^8$ ions pass through the ion channels per second, creating a localized ion gradient field near the ion channel. This localized ion diffusion field alters the local refractive index of the medium and acts as an "ion nanolens", generating an interferometric scattering signal (Fig. 1b). Finite element simulations show that this ion diffusion field has a characteristic scale of ~183 nm (Fig. 1c), and the integral ion concentration within different ranges of the diffusion field was tracked over time (see SI section 9 for details). Thus, in principle, the local ion diffusion field could provide a significant EM-iSCAT signal.

In imaging experiments, the time series of raw iSCAT images contain the spatiotemporal information about the ion field as well as the cell structure. Then the raw iSCAT images were processed via a Short-Time Fourier Transform (STFT) program pixel to pixel to reconstruct the EM-iSCAT images (Fig. 1d). This transform extracted the amplitude only at the modulation frequency and filtered out the noise at other frequency (Fig. 1e). Since the modulation did not change the cell structure, the final EM-iSCAT images excluded the structural characteristics of cells, and only obtained charge density distributions on cell membrane. Similar to optical electrochemical impedance microscopy[14,29], the amplitude ($\Phi$) of the EM-iSCAT images is proportional to the localized charge density, i.e. the ion diffusion field (see SI section 3 for details) under a given modulation. Therefore, this EM-iSCAT approach could image the charge density on cell membranes, or even the ion field originated from single ion channels.

Analogical to patch-clamp[2], our EM-iSCAT microscopy supports two operation modes: the whole-cell mode and the single-channel mode. In the whole-cell mode, the cellular electrophysiological analysis was achieved by tracking the overalls change of charge density across the whole cell membrane (Fig. 1f). In the single-channel mode, individual ion channels are localized and tracked, thereby the "open and close" activities of each single ion channel or clustered channels were recorded (Fig. 1g). The temporal resolution of the single-channel mode was mainly limited by the camera speed and the modulation frequency. Our current instrumental configuration realized a 650 μs temporal resolution. This value is essentially comparable to an early patch-clamp, allowing the study of ion channel activity at the microsecond scale. For faster processes related to ion channels, the temporal resolution of EM-iSCAT could be further improved in principle by upgrading faster cameras and increasing modulation frequencies.

**Whole-cell EM-iSCAT electrophysiology**

First, we show that the amplitude ($\Phi$) of EM-iSCAT is indeed related to the ion channel activity of living cells. As is known, a variety of ion channels are distributed on cell membranes and help maintain normal cell functions. The EM-iSCAT of living cells show a strong signal amplitude of ~5.6 compared to the background of ~0.8 from ITO (Fig. 2a). When cells were fixed by 4% paraformaldehyde (PFA), membrane proteins associated with ion transport on the cell membrane were destroyed, so that the signal drops significantly within a few seconds. We noted that the EM-iSCAT images of living cells exhibits discretely distributed bright spots rather than a uniform distribution (Fig. 2b and Supplementary Video 1). We identified these bright spots as "active zones", and these active regions show a signal ~4.5 times stronger than fixed cells or ITO background (Fig. 2c). This clearly shows that the ion field on the cell membrane is not uniform, but concentrated in the localized region surrounding the ion channel, which is consistent with the simulation results. Once cells were fixed, the "active zones" in the EM-iSCAT images almost disappeared, with the overall signal reduction (Fig. 2d). These proof-of-principle experiments proved the viability of measuring the local ion density on the cell membrane via the signal amplitude and image of EM-iSCAT.

We then used the whole-cell analysis mode to disclose the cellular response to osmotic pressure changes in the extracellular environment. As shown in Fig. 2e, cells could adjust osmotic balance by regulating ion transport (mainly $K^+$, $Na^+$) and water exchange across cell membranes[30-32]. We added sodium-free and high-sodium medium to the sample chamber to change the osmotic pressure in the extracellular environment and then monitored the cellular response (see SI section 4 for details). In three cycles of hypotonic stimulation (Fig. 2f), the differential signal of whole cell, $\Delta\Phi = \Phi_i - \Phi_0$ (where $\Phi_0$ is the whole cell signal at t = 0), periodically rises and falls. When the external osmotic pressure suddenly decreases, the cell rapidly activates the outward expulsion of ions, resulting in charge density increase at the outer interface of the cell membrane. The EM-iSCAT images also show that during the response state (Fig. 2g, **2**), the "active zones" on cell membrane is greatly increased compared to the resting state (Fig. 2g, **1**). In three cycles of hypertonic stimulations (Fig. 2h), the charge density at the outer interface of cell membrane rapidly decreases due to the inward transport of ions, and the overall differential signal on the cells periodically decreases and recovers. For illustrative purposes, we plotted the EM-iSCAT amplitude image of hypertonic stimulations with the absolute value of the signal change $|\Delta\Phi|$. Similar to hypotonic stimulation, the "active zones" in the response state (Fig. 2i, **4**) is also significantly increased compared to the resting state (Fig. 2i, **3**). The process for "active zones" increasing in osmotic changing are presented in Supplementary Video 2-3.

To verify that the EM-iSCAT signal exactly originates from the ion transport across the ion channels, we blocked the $K^+$ channels by using three different nonselective $K^+$ channel blockers, including quinine (1 mM), 4-aminopyridine (4-AP, 1 mM), and tetraethylammonium (TEA, 1 mM). Under the same hypotonic stimulation, cells treated by these blockers no longer responded to the osmotic pressure (Fig. 2j). In addition, the "active zones" almost did not change after hypotonic stimulation in the presence of the blockers (Fig. 2k). Further, by comparing the images during the response and resting state in one hypertonic stimulation cycle, a small area of the "active zone" (Fig. 2g, **2**, green box) and another "inactive zone" (Fig. 2g, **2**, blue box) were selected to extract the signal trajectories. We found that signals in the "active zone" show multiple peaks while the "inactive zone" is close to the background (Fig. 2l). Therefore, we demonstrate that these bright spots in the "active zones" are related to the activity of ion channels. Some spots may be from individual channels, while others may be ascribed to clustered channels. Due to the optical diffraction limit, we cannot directly resolve the details of individual ion channels in real space. However, the activity of single ion channels on the cell membrane can still be monitored when they are far away from each other.

**Single-channel EM-iSCAT electrophysiology**

NMDARs are ligand-gated ion channels in the central nervous system (CNS), which play key physiological roles in synaptic function, such as synaptic plasticity, learning, and memory[33]. When NMDARs are activated and turned on, positively charged ions, such as $Na^+$ and $K^+$, pass through the cell membrane and form a diffusion field near the ion channel port (Fig. 3a). Here we demonstrated the feasibility of EM-iSCAT to monitor the activity of single ion channels using NMDARs. For this experiment, human GluN1 and GluN2A cDNAs were transfected into HEK293 cells, and the level of NMDAR expression was determined by confocal fluorescence imaging (Fig. S1). To activate NMDARs, 0.1 mM glycine and 1 mM glutamic acid were added to the extracellular medium, and the whole-cell EM-iSCAT signal is shown in Fig. 3b. Compared with the control group without transfection, the signal of NMDAR-expressing cells increased in the presence of agonists, which corresponds to the charge density change on the cell membrane caused by transmembrane ions transport through NMDARs. We further monitored the EM-iSCAT signal of single ion channels and tracked their movement trajectories on the membrane in real time (Fig.3c, see method, SI section 7 and Supplementary Video 4 for details). Due to the cell membrane fluidity, ion channel proteins are restricted to sub-diffusion within a certain spatial range[34-36]. We located the position of ion channels via a two-dimensional (2D) Gaussian fitting of the bright spots in the sequential EM-iSCAT images (snaps in Fig. 3c), and delineated the tracks using the central coordinates. The median diffusion coefficient of this channel on cell membrane was calculated as $411 \times 10^{-4}$ μm$^2$/s, which matched previous reports well[37,38]. Meanwhile, the tracking trajectory ensures that the real-time signals we obtained are from the same ion channel. Three cases of single channel signals are shown in Fig. 3c, which present typical turnover features with two states of "on" and "off". The signal intensity was determined by $\Delta A = A_i - A_0$, where $A_i$ is the amplitude of the 2D Gaussian fitting function and $A_0$ is the mean intensity of EM-iSCAT snaps at "off" state within the first few milliseconds of recording. Figure 3d shows a zoom-in view of a segment of Ch 3, the presence and absence of the bright "active zone" indicate the open and close states of the channel, respectively, and the temporal resolution of single channel mode is 650 μs (Fig. 3d, inset).

We counted the intensity histogram and the open duration time of all single-channel signals on cell membrane (Fig. 3e, f). The histogram presents two clear peaks: $\Delta A = 0.07$ and $\Delta A = 7.34$, corresponding to the completely close and open state of NMDARs, respectively. The histogram of dwell time was extracted from the single channel trajectories (Fig. 3c and Fig. S10a). The "on" dwell time show two components of 2.89 ms (59.14%) and 24.43 ms (40.86%), and the "off" dwell time show three components of 3.01 ms (51.4%), 35.23 ms (33.25%) and 316.48 ms (15.35%) (Fig. S10c). These results were consistent with previous measurements of NMDARs by patch-clamp[39-41]. Due to limitations of current imaging speed (1.5 kHz), we can only track open events with open duration above millisecond for now, but the use of high-speed cameras will allow EM-iSCAT to achieve higher temporal resolution. We exponentially fitted the rise time ($t_{up}$) and the fall time ($t_{down}$) of individual turn over events of single channels (Fig. 3g), and the statistical results were shown in Fig. 3h. It should be noted that the gating kinetics of NMDARs depend on the agonist binding steps and conformational transformation of receptor protein[42]. In patch-clamp measurement, the detected ion current is derived from a real ion stream through the ion channel. Generally, the explosive opening steps after pre-opening gating steps is much faster than the instrument response time, thus the temporal resolution of patch-clamp mainly relies on the bandwidth of electronic circuits as well as the data acquisition speed.[33,43]

In EM-iSCAT, the signal is from the ion diffusion field nearby the ion channel port, the response time is mainly limited by the time scale of the formation ($t_{up}$) and dissipation ($t_{down}$) of the ion diffusion field. When the ion channel is opening and closing, the distributions of $t_{up}$ and $t_{down}$ are similar, both with an averaged value around 0.17 ms. This value is close to the simulation results calculated at 200-250 nm away from the ion channel (see Table S1 and SI section 9 for details). In principle, the response time of EM-iSCAT imaging could be down to ~170 μs, corresponding to 5-10 k bandwidth, which is sufficient for most ion channel studies.

In addition to above single-channel characteristics, we also found some dual-channel signals (Fig. 3i, j). The signal intensity histogram of dual channels shows three separated peaks, $\Delta A$ = 0.09, 7.53 and 14.95, corresponding to the double close, single open and double open states, respectively (Fig. 3k). As mentioned above, when two channels are close enough[44], forming a dimer or at a distance within the diffraction limit, they still appear as one bright spot on the EM-iSCAT image. However, from the characteristics of the turnover signals and the histogram peaks, we can clearly distinguish the behavior of the single or dual channels, or even more channels clustered together. The simulation results also confirmed our experimental observation (Fig. S14 and Supplementary Video 5-6).

**Identifying channel types and cellular response**

With both global imaging and single-channel resolution, EM-iSCAT has the potential to open up new avenues of electrophysiological studies that are difficult to achieve in the past. For example, it is challenging to directly locate and identify different kinds of ion channels on cell membrane. It usually requires fluorescence labeling, and the labels may affect the normal physiological function of cells. Here, we distinguished different types of ion channels by their different response modes and spatial locations. Variations in extracellular cation concentration can alter the concentration gradient of corresponding ions inside and outside the cell membrane, triggering activities of related ion channels. We increased the concentration of one specific cation ($K^+$, $Na^+$, and $Ca^{2+}$, respectively) while keeping the same extracellular osmotic pressure (see SI section 8 for details) to trigger specific channels, and extracted the "active regions" on the same cell (Fig. 4a). Cells exhibit different whole-cell signals in each cation environment (Fig. 4b), channel distributions under different cation stimulations do not overlap.

In the "active regions" of the images, we observed single-channel behaviors (Fig. 4b inset), which show different amplitude intensities in accordance with whole-cell responses. The spatial positions of these active regions under different cation environment are basically separated (Fig. 4a), and the signal features of ion channels found in these regions are also distinguishable (Fig. 4c). These results suggest that there are different kinds of ion channels responding to different cations, and they tend to be located at different sites on the cell membrane. We also noticed partial overlap in spatial locations and response intensities, it may be attributed to some common regulatory pathways that respond to different cationic environment. Unlike patch-clamp that distinguishes different kinds of ion channels by their conductance, EM-iSCAT resolves different channels based on a more diverse set of physical characteristics of ions themselves, such as their distinct dielectric constant and mobility in medium (Fig. S15).

In addition to identifying different ion channels, the EM-iSCAT method can also directly resolve single channels or channel clusters. Nicotinic acetylcholine receptors (nAChRs) are a class of ligand-gated non-selective cationic channels widely expressed in CNS. nAChRs function as neurotransmitter receptors that respond to endogenous acetylcholine and choline, modulating neuronal excitability and synaptic communication. We treated human neuroblastoma cells (SH-SY5Y) that naturally expressing several subtypes of nAChR[45,46], such as α7 and α3β4, with nicotine (Fig. 4d). After stimulation, we observed the behaviors of single channels and several channels aggregated as channel cluster from EM-iSCAT images (Fig. 4e). Clustering is the default mode of organization and working of ion channels in the cell membrane[47], and clustering of nAChRs is essential for their proper function and localization within the synapse, such as the neurotransmission at neuromuscular junctions[48,49]. Our EM-iSCAT microscopy provides a convenient way to study ion channel clustering and interactions, with both spatial resolution and single channel sensitivity.

We further explored a new research paradigm to decrypt the spatio-temporal heterogeneity of cellular response in a cell community. Five cells under hypotonic stimulation were imaged with EM-iSCAT simultaneously (Fig.4f), and the whole-cell signals of these five cells were extracted respectively over time (Fig. 4g). Different cells showed heterogeneity in both response intensity and dynamics, which might be caused by various number and activity of related ion channels on different cells. All in all, EM-iSCAT allows the localization and distinguishing of different channels or channel clusters, and measurement of heterogeneous responses in cell community, which is expected to provide a new research tool for revealing signaling mechanisms in neural networks.

**Discussion**

Label-free optical electrophysiology has emerged as a new area for the study of electrical activities of cells and neural networks *in vitro*. The advantages including non-perturbative nature, flexible spatial selectivity, long-term imaging capabilities, and multi-channel measurement make it an active field full of possibilities. In this work, we have developed an EM-iSCAT imaging method that is capable of monitoring the real-time cellular electrical responses and enabling label-free optical imaging of single ion channel activity for the first time.

We first used EM-iSCAT to measure ion transmembrane transport processes at the whole-cell level by analyzing the changes in cell's optical signals during cell inactivation and osmotic regulation. Subsequently, we detected the opening and closing processes of single NMDARs in response to glutamic acid and glycine activation with microsecond time resolution. The gating dynamics results are consistent with those obtained in the patch-clamp experiments. More importantly, we are able to distinguish and localize different ion channels on the cell membrane, identify channel clusters and realize the simultaneous detection of multiple cells. Although the current precision of localization is still within the diffraction limit, it may be possible to achieve nanometer accuracy in combination with super-resolution techniques[50]. Although in this work we used ligand-gated ion channels, the optical imaging methods can be combined in principle with electrical stimulation techniques or even photothermic regulation techniques in order to study voltage-gated or temperature-gated ion channels ion channels.

We remark that this method is easy to use for neurobiologists. Firstly, compared with voltage-sensitive fluorescence imaging, EM-iSCAT requires no labeling, which can minimize the effect on ion channels and the damage to cells; Secondly, compared with patch-clamp, EM-iSCAT extracts signals directly from the image and does not require special operation training, such as cell sealing. In addition, despite still being a lab-made setup, EM-iSCAT could be an integrated instrument with automatic data acquisition and analysis. It is also possible to achieve AI-assistant channel identification by building a database of signaling characteristics of different ion channels and combing it with machine learning. Ultimately, this technology will provide new research tools for a wide range of neuroscience related to cell membrane potentials and ion channels.

**References:**

1    Neher, E. & Sakmann, B. Single-channel currents recorded from membrane of denervated frog muscle-fibers. *Nature* **260**, 799-802 (1976).

2    Sakmann, B. & Neher, E. Patch clamp techniques for studying ionic channels in excitable-membranes. *Annu. Rev. Physiol.* **46**, 455-472 (1984).

3    Du, H. *et al.* The cation channel TMEM63B is an osmosensor required for hearing. *Cell Rep.* **31** (2020).

4    Peralta, F. A. *et al.* Optical control of PIEZO1 channels. *Nat. Commun.* **14** (2023).

5    Siegel, M. S. & Isacoff, E. Y. A genetically encoded optical probe of membrane voltage. *Neuron* **19**, 735-741 (1997).

6    Stosiek, C., Garaschuk, O., Holthoff, K. & Konnerth, A. In vivo two-photon

calcium imaging of neuronal networks. *Proc. Natl Acad. Sci. USA* **100**, 7319-7324 (2003).

7   Kralj, J. M., Hochbaum, D. R., Douglass, A. D. & Cohen, A. E. Electrical spiking in escherichia coli probed with a fluorescent voltage-indicating protein. *Science* **333**, 345-348 (2011).

8   Peterka, D. S., Takahashi, H. & Yuste, R. Imaging voltage in neurons. *Neuron* **69**, 9-21 (2011).

9   Scanziani, M. & Hausser, M. Electrophysiology in the age of light. *Nature* **461**, 930-939 (2009).

10  Ionescu-Zanetti, C. *et al.* Mammalian electrophysiology on a microfluidic platform. *Proc. Natl Acad. Sci. USA* **102**, 9112-9117 (2005).

11  Finkel, A. *et al.* Population patch clamp improves data consistency and success rates in the measurement of ionic currents. *J. Biomol. Screening* **11**, 488-496 (2006).

12  Jones, I. L. *et al.* The potential of microelectrode arrays and microelectronics for biomedical research and diagnostics. *Anal. BioanalChem.* **399**, 2313-2329 (2011).

13  Spira, M. E. & Hai, A. Multi-electrode array technologies for neuroscience and cardiology. *Nat. Nanotechnol.* **8**, 83-94 (2013).

14  Liu, X. W. *et al.* Plasmonic-based electrochemical impedance imaging of electrical activities in single cells. *Angew. Chem. Int. Ed.* **56**, 8855-8859 (2017).

15  Yang, Y. Z. *et al.* Imaging action potential in single mammalian neurons by tracking the accompanying sub-nanometer mechanical motion. *ACS Nano* **12**, 4186-4193 (2018).

16  Habib, A. *et al.* Electro-plasmonic nanoantenna: A nonfluorescent optical probe for ultrasensitive label-free detection of electrophysiological signals. *Sci. Adv.* **5** (2019).

17  Alfonso, F. S. *et al.* Label-free optical detection of bioelectric potentials using electrochromic thin films. *Proc. Natl Acad. Sci. USA* **117**, 17260-17268 (2020).

18  Ling, T. *et al.* High-speed interferometric imaging reveals dynamics of neuronal deformation during the action potential. *Proc. Natl Acad. Sci. USA* **117**, 10278-10285 (2020).

19  Kukura, P. *et al.* High-speed nanoscopic tracking of the position and orientation of a single virus. *Nat. Methods* **6**, 923-U985 (2009).

20  Park, J. S. *et al.* Label-free and live cell imaging by interferometric scattering microscopy. *Chem. Sci.* **9**, 2690-2697 (2018).

21  Lindfors, K., Kalkbrenner, T., Stoller, P. & Sandoghdar, V. Detection and spectroscopy of gold nanoparticles using supercontinuum white light confocal microscopy. *Phys. Rev. Lett.* **93** (2004).

22  Ortega-Arroyo, J. & Kukura, P. Interferometric scattering microscopy (iSCAT): new frontiers in ultrafast and ultrasensitive optical microscopy. *Phys. Chem. Chem. Phys.* **14**, 15625-15636 (2012).

23  Cole, D., Young, G., Weigel, A., Sebesta, A. & Kukura, P. Label-free single-molecule imaging with numerical-aperture shaped interferometric scattering microscopy. *ACS Photonics* **4**, 211-216 (2017).

24  Taylor, R. W. *et al.* Interferometric scattering microscopy reveals microsecond nanoscopic protein motion on a live cell membrane. *Nat. Photonics* **13**, 480-487 (2019).

25  Park, J.-S. *et al.* Fluorescence-combined interferometric scattering imaging reveals nanoscale dynamic events of single nascent adhesions in living cells. *J. Phys. Chem. Lett.* **11**, 10233-10241 (2020).

26  Merryweather, A. J., Schnedermann, C., Jacquet, Q., Grey, C. P. & Rao, A. Operando optical tracking of single-particle ion dynamics in batteries. *Nature* **594**, 522-528 (2021).

27  Lyu, P.-T. *et al.* Decrypting material performance by wide-field femtosecond interferometric imaging of energy carrier evolution. *J. Am. Chem. Soc.* **144**, 13928-13937 (2022).

28  Merryweather, A. J. *et al.* Operando monitoring of single-particle kinetic state-of-charge heterogeneities and cracking in high-rate Li-ion anodes. *Nat. Mater.* **21**, 1306-1313 (2022).

29  Wang, W. *et al.* Single cells and intracellular processes studied by a plasmonic-based electrochemical impedance microscopy. *Nat. Chem.* **3**, 249-255 (2011).

30  Lang, F. *et al.* Functional significance of cell volume regulatory mechanisms. *Physiol. Rev.* **78**, 247-306 (1998).

31  Wehner, F. Cell volume-regulated cation channels. *Cell Volume Regulation* **123**, 8-20 (1998).

32  Gillespie, P. G. & Walker, R. G. Molecular basis of mechanosensory transduction. *Nature* **413**, 194-202 (2001).

33  Hansen, K. B. *et al.* Structure, function, and allosteric modulation of NMDA receptors. *J. Gen. Physiol.* **150**, 1081-1105 (2018).

34  Dahan, M. *et al.* Diffusion dynamics of glycine receptors revealed by single-quantum dot tracking. *Science* **302**, 442-445 (2003).

35  Hoze, N. *et al.* Heterogeneity of AMPA receptor trafficking and molecular interactions revealed by superresolution analysis of live cell imaging. *Proc. Natl Acad. Sci. USA* **109**, 17052-17057 (2012).

36  Lippert, A. *et al.* Single-molecule imaging of Wnt3A protein diffusion on living cell membranes. *Biophys. J.* **113**, 2762-2767 (2017).

37  Groc, L. *et al.* NMDA receptor surface mobility depends on NR2A-2B subunits. *Proc. Natl Acad. Sci. USA* **103**, 18769-18774 (2006).

38  Yadav, R. & Lu, H. P. Probing dynamic heterogeneity in aggregated ion channels in live cells. *J. Phys. Chem. C* **122**, 13716-13723 (2018).

39  Popescu, G. & Auerbach, A. Modal gating of NMDA receptors and the shape of their synaptic response. *Nat. Neurosci.* **6**, 476-483 (2003).

40  Popescu, G., Robert, A., Howe, J. R. & Auerbach, A. Reaction mechanism determines NMDA receptor response to repetitive stimulation. *Nature* **430**, 790-793 (2004).

41  Sasmal, D. K., Yadav, R. & Lu, H. P. Single-molecule patch-clamp FRET anisotropy microscopy studies of NMDA receptor ion channel activation and deactivation under agonist ligand binding in living cells. *J. Am. Chem. Soc.* **138**, 8789-8801 (2016).

42    Kussius, C. L. & Popescu, G. K. Kinetic basis of partial agonism at NMDA receptors. *Nat. Neurosci.* **12**, 1114-1120 (2009).

43    Amin, J. B., Gochman, A., He, M. M., Certain, N. & Wollmuth, L. P. NMDA receptors require multiple pre-opening gating steps for efficient synaptic activity. *Neuron* **109**, 488-501 (2021).

44    Yadav, R. & Lu, H. P. Revealing dynamically-organized receptor ion channel clusters in live cells by a correlated electric recording and super-resolution single-molecule imaging approach. *Phys. Chem. Chem. Phys.* **20**, 8088-8098 (2018).

45    Lukas, R. J. & Eisenhour, C. M. Interactions between tachykinins and diverse, human nicotinic acetylcholine receptor subtypes. *Neurochem. Res.* **21**, 1245-1257 (1996).

46    Canastar, A. *et al.* Promoter methylation and tissue-specific transcription of the α7 nicotinic receptor gene, CHRNA7. *J. Mol. Neurosci.* **47**, 389-400 (2012).

47    Dixon, R. E., Navedo, M. F., Binder, M. D. & Santana, L. F. Mechanisms and physiological implications of cooperative gating of clustered ion channels. *Physiol. Rev.* **102**, 1159-1210 (2022).

48    Huk, K. H. & Fuhrer, C. Clustering of nicotinic acetylcholine receptors: From the neuromuscular junction to interneuronal synapses. *Mol. Neurobiol.* **25**, 79-112 (2002).

49    Hogg, R. C., Raggenbass, M. & Bertrand, D. *Nicotinic Acetylcholine Receptors: from Structure to Brain Function*, in: Reviews of Physiology, Biochemistry and Pharmacology, vol 147. Springer, Berlin, Heidelberg (2003).

50    Kuppers, M., Albrecht, D., Kashkanova, A. D., Luhr, J. & Sandoghdar, V. Confocal interferometric scattering microscopy reveals 3D nanoscopicstructure and dynamics in live cells. *Nat. Commun.* **14**, 1962-1962 (2023).

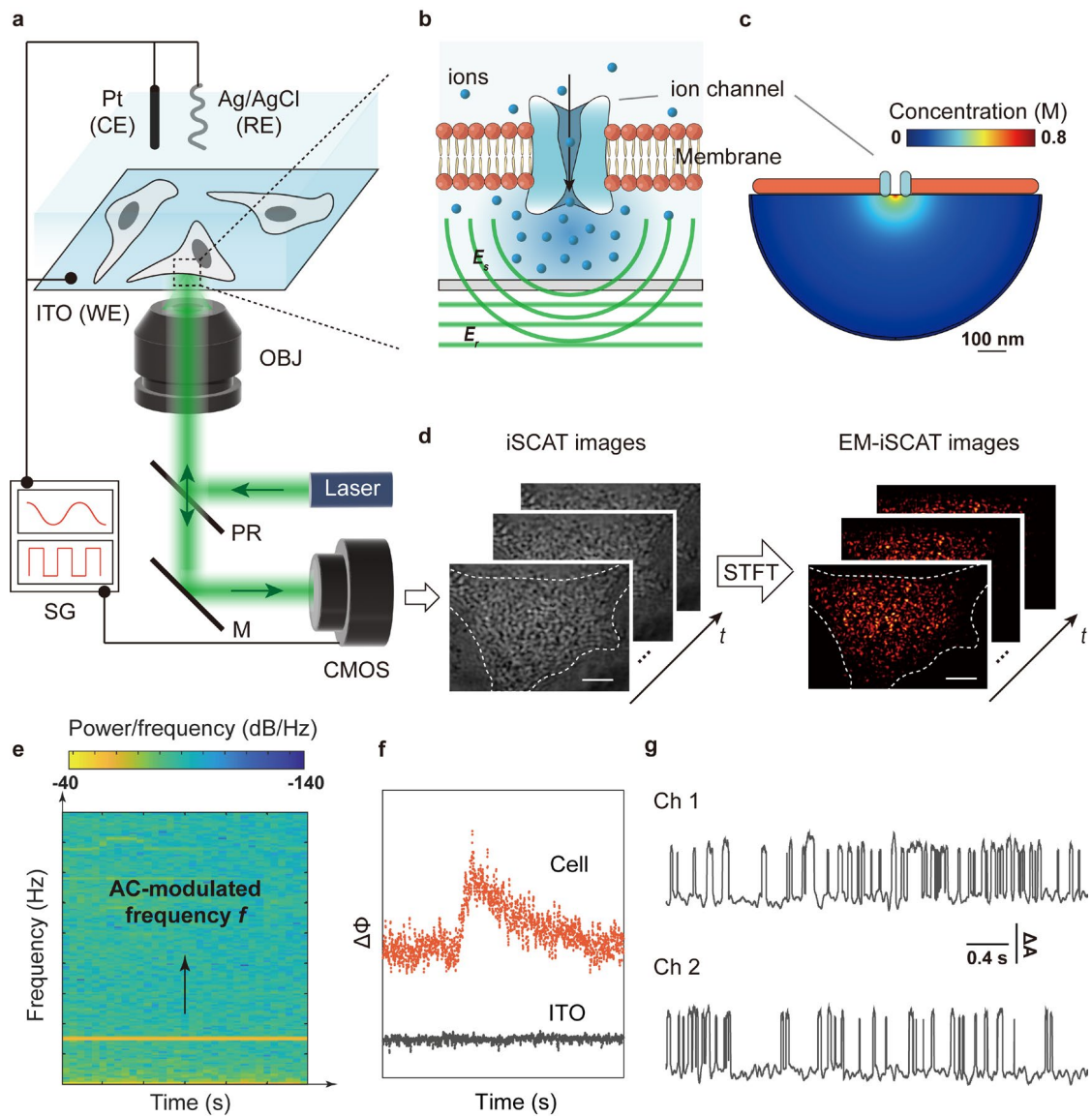

**Fig. 1 | Concept of EM-iSCAT microscopy. a,** Experimental setup of EM-iSCAT microscopy for ion channel activity imaging. Cells are plated on a ITO conductive glass, a digital signal generator (SG) is used to trigger and synchronize a sinusoidal voltage modulation and iSCAT imaging. OBJ, 60× objective; PR, partial reflector; M, mirror. **b,** The iSCAT signal from ions that pass through ion channels and form a local diffusion filed. **c,** The theoretical model of the ion diffusion field. **d,** Time series of EM-iSCAT images are obtained by applying STFT analysis to the iSCAT image sequence. The white dashed line gives the cellular outline based on the bright-field image. Scale bars: 5 μm. **e,** Power spectrum using STFT. **f,** The STFT amplitude change (ΔΦ) on the whole cell and ITO background, respectively. **g,** Two single channel signals (ΔA) extracted from EM-iSCAT images of HEK-293 cells expressing NMDARs.

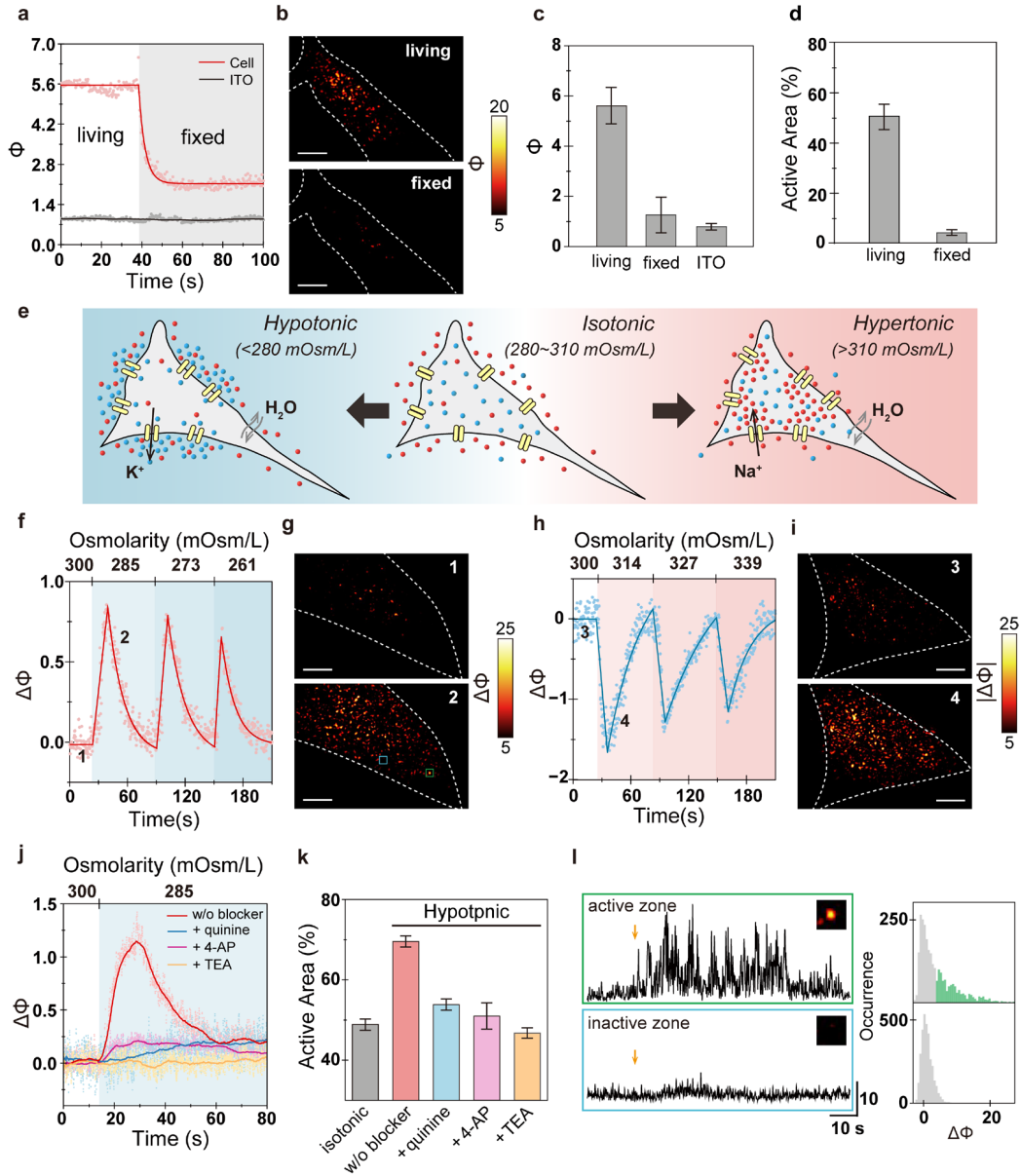

**Fig. 2 | Whole-cell electrophysiological activity analysis. a,** The changes of the mean STFT amplitude (Φ) on the whole cell and ITO during the fixation process with 4% PFA, respectively. **b,** EM-iSCAT images of the living cell and the fixed cell. **c,** Φ of living cells, fixed cells and ITO, respectively; n = 5 cells. **d,** The proportion (%), both in living cells and fixed cells, of the active area to the total cell area in the EM-iSCAT images; n = 5 cells. **e,** Schematic of the behavior of ion channels on cell membranes in isotonic, hypotonic and hypertonic environments. **f,** Whole-cell signals (ΔΦ) on the same cell during three hypotonic stimulation processes. **g,** EM-iSCAT images of cellular response in initial isotonic environment **1** and after hypotonic stimulation **2**. **h,** Whole-cell signals (ΔΦ) on the same cell during three hypertonic stimulation processes.

**i,** EM-iSCAT images of cellular response in initial isotonic environment **3** and after hypertonic simulation **4**. **j,** $\Delta\Phi$ of one hypotonic stimulation process on cells treated with/without three different $K^+$ channel blockers. **k,** The proportion of "active zones" under the isotonic environment and hypotonic simulation after being treated with/without blockers; n = 4-6 cells. **l,** $\Delta\Phi$ versus times (left panel) for the "active zone" (green) and "inactive zone" (blue) extracted from the EM-iSCAT image in hypotonic simulation (orange arrow), and histograms (right panel), respectively. Scale bars in b, g, and i: 5 μm.

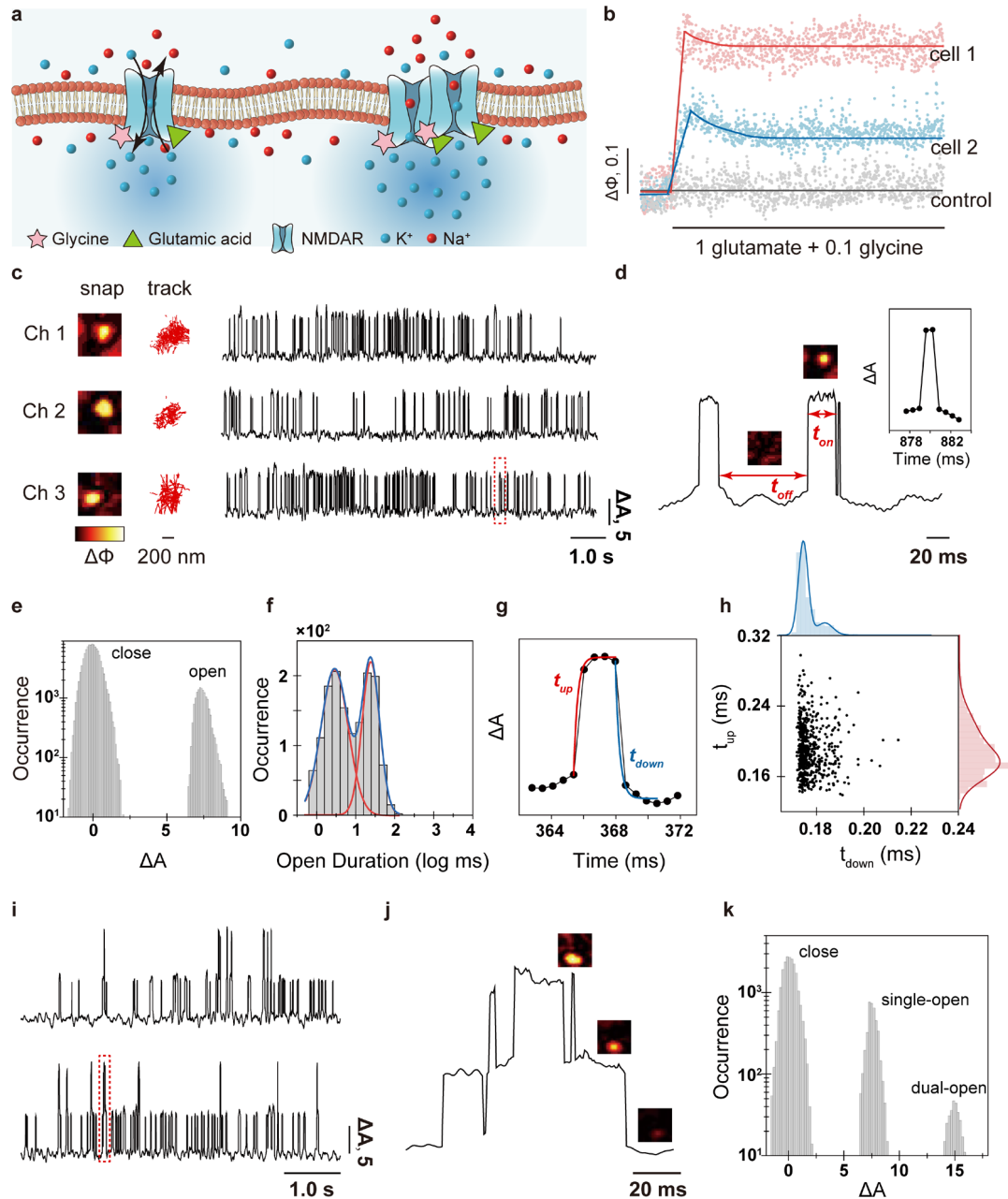

**Fig. 3 | Detection of single NMDAR signal. a,** Schematic of ion transport through NMDARs. **b,** Whole-cell signals of two transfected HEK-293 cells (cell 1 and cell 2) and a non-transfected cell (control) after glutamic acid + glycine stimulation. **c,** Single channel signals from NMDARs in HEK-293 cells. Snaps of some opening events and the tracks of the channels are listed beside the signal trajectories. **d,** A zoomed-in view of a segment of Ch 3 (red dashed box), $t_{on}$ and $t_{off}$ represent dwell time of channel opening and closing, respectively. **e,** Histograms of signal amplitude, derived from the single channel trajectories in (c) and Fig. S8(a). **f,** Histograms of time components of NMDARs calculated from all the trajectories. The blue line is for overall distribution,

and the red line is for individual time components. **g,** One opening event of a single ion channel, the rise time ($t_{up}$, red curve) and fall time ($t_{down}$, blue curve) are obtained by first order of kinetic fitting. **h,** The scatter plot and the histogram show the fitting results of $t_{up}$ and $t_{down}$ respectively. **i,** Representative trajectories of dual channel events. **j,** A zoom-in view with snaps of a segment of the trajectory (red dashed box) in (i). **k,** Histograms of signal amplitude derived from (i).

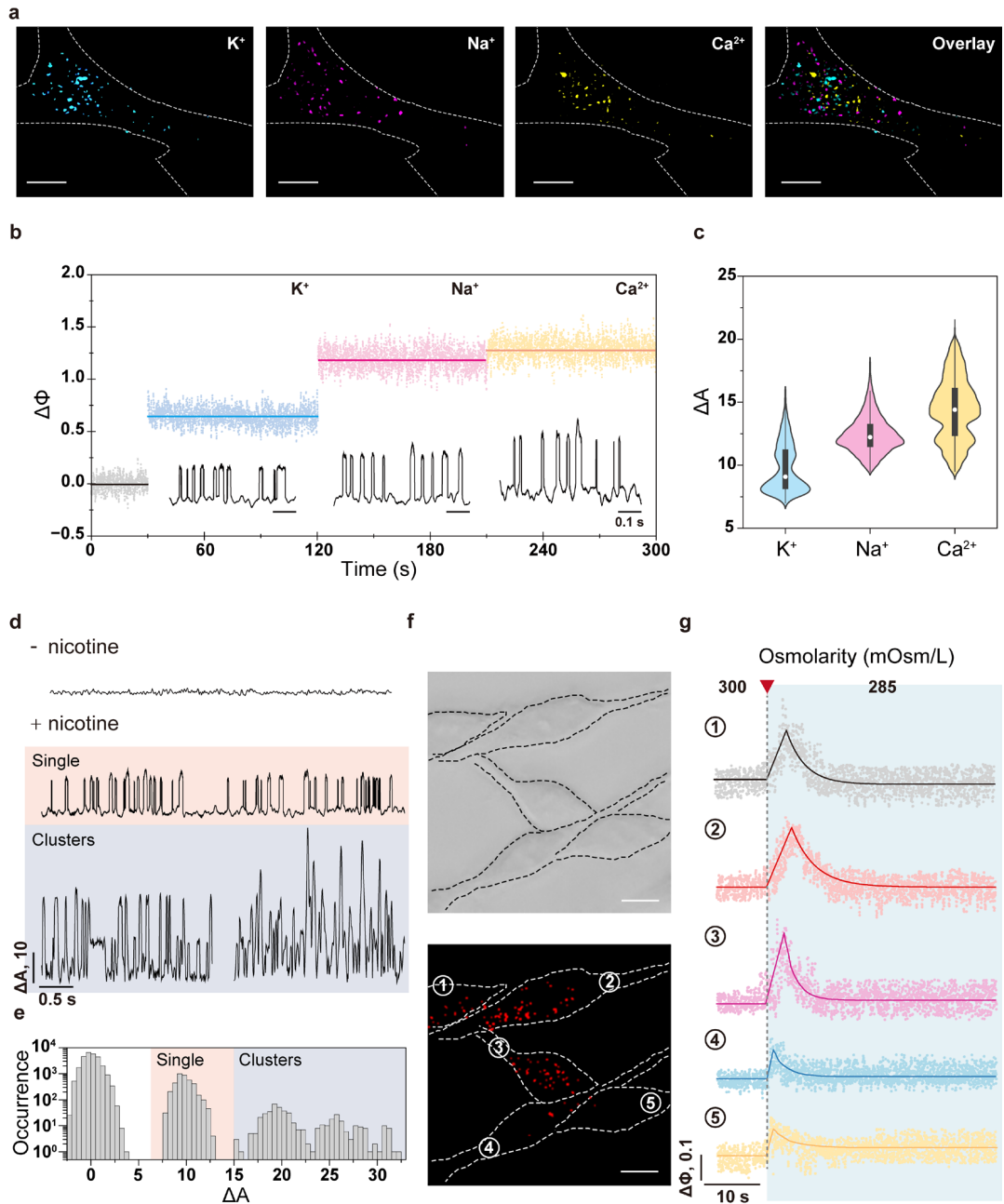

**Fig. 4 | Single-channel localization and cell community responses imaging. a,** Maps of the distribution of single-channel activity in different cationic environments. **b,** Whole-cell signals in high $K^+$, $Na^+$ and $Ca^{2+}$ concentration environments, respectively. Inset, several representative segments of single-channel activity trajectories in each case. **c,** Statistics intensity single-channel activity in different environments. **d,** nAChRs signals obtained after stimulation of SH-SY5Y cells with/without nicotine (100 μM). Signals of single (orange) and multiple (blue) nAChRs were extracted from EM-iSCAT images. **e,** The histogram of amplitudes for all signals shown in (d). **f,** EM-iSCAT

imaging of multicellular networks. Top panel, bright field image of a multicellular network of five cells. Bottom panel, EM-iSCAT images of cellular responses after hypotonic stimulation. **g,** Whole-cell signals of the five cells shown in (d) during hypotonic stimulation. Scale bar: a, 5 μm; f, 10 μm.

## Methods

### Sample preparation

**Cell culture.** PC-12 cells were cultured in DMEM medium supplemented with 10% fetal bovine serum (FBS), HEK-293 cells were cultured in MEM medium supplemented with 10% FBS, SH-SY5Y cells were cultured in MEM/F12 medium supplemented with 15% FBS. All cells were kept in a humidified atmosphere at 37 ºC and 5% $CO_2$ / 95% air. For measurement, the cells were plated onto ITO electrode and grown to 80% confluency. Before experiments, the culture medium was discarded and cells were rinsed twice with 200 μl phosphate buffer saline (PBS).

**Preparation of electrochemical cell.** ITO-coated glass slides (coating thickness, ~23 nm; resistance, 80-100 Ω/square) were cleaned by sonication sequentially in acetone, ethanol, and DI water, each for 20 min, and then dried with $N_2$. A 3 mm thick polydimethylsiloxane (PDMS) with a circular hole of 10 mm at its center was attached to ITO surface, assembled into a culture chamber for cells. Both ITO slides and PDMS were irradiated under UV for 30 min before cell seeding.

**Expression of NMDAR in HEK-293 cell.** HEK-293 cells were cultured in a 10 mm PDMS chamber in MEM supplemented with 10% FBS at 37 ºC in a 5% $CO_2$ atmosphere. Gene expression vectors encoding Human GluN1 (pCAG-mNR1-GFP) and GluN2A (pCAG-mN2A-mcherry) were purchased from Vigenebio (Shandong, China). Vectors were used for expression without further amplification or purification. GluN1 and GluN2A victors (0.5 μg: 0.5 μg) were transfected into the HEK-293 cells, when the cell confluence was about 80%. Lipofectamine 3000 (Invitrogen, Carlsbad, CA) transfection reagent was used according to the manufacturer's protocol. The expression of NMDAR in HEK-293 cells was examined by confocal fluorescence imaging (TCS SP8 STED, Leica, Fig.S1), immediately after EM-iSCAT detection and fixation in 4% PFA in PBS.

### Optical setup

The experimental setup is depicted schematically in Supplementary Figure S2. The imaging setup was built on an inverted microscope (Nikon Ti2-U). The collimated output of a 532 nm laser (MW-GLN-532, Changchun Laser Optoelectronics

Technology Co., Ltd.) is passed through a 5× achromatic Galilean beam expander (GBE05-A, Thorlabs). A 300 mm focal length len focuses the beam onto the back focal plane of the microscope objective (S Plan Fluor ELWD 60×/0.7 NA, Nikon). The signal collected by the objective is transmitted through a partial reflector (PR) and focused onto a sCMOS camera (Dhyana 400D, Tucsen). The PR consists of a 30 nm thick circular gold layer of 2 mm diameter (transmissivity of 10%) evaporated onto a 10:90 (R:T) plate beam splitter. An AC-modulated voltage was applied to the ITO slide with a potentiostat (CH150, Corrtest) using a three-electrode configuration, where the ITO, a Ag/AgCl wire and a Pt coil served as the working, reference and counter electrodes, respectively. The size of the working electrode and the distance between working and counter electrodes were maintained by a double-sided tape (thickness: 89μm, 3M Tape 665) with a 2 mm hole. A digital signal generator (DG1000Z, RIGOL) was used to synchronize the applied potential and image acquisition. All data were recorded at 25 °C by a home-built sample stage heater.

**Data acquisition**

For the experiment presented in Fig. 2 and 4, images were acquired with an exposure time of 300 μs, at a frame rate of 300 Hz, and the voltage modulation frequency was set to 30 Hz, with an amplitude of 1V. Experiments presented in Fig. 3 were carried out using the same exposure time at a frame rate of 1.5 kHz, and the voltage modulation frequency was set to 150 Hz, with an amplitude of 1.2 V.

**Data analysis**

Unless otherwise stated, analysis was performed using custom written code in MATLAB (MathWorks). To remove the static scattering background from the ITO surface and cell membrane structure, ratiometric images, $I$, were calculated as $I = (I_i - I_{bg})/I_{bg}$, where $I_{bg}$ is the temporal median intensity background calculated from first 300 frames of image sequence. We performed a STFT analysis on each pixel of ratiometric image sequence with a Hann window length of 128 frames and rolling step length of 1 frame. Then we extracted the amplitude averaged over 1 s, and constructed the STFT amplitude image (EM-iSCAT image).

For whole-cell detection (Fig. S3a), the Hann window's overlap length was set to 64 frames, and the mean intensity within the profile of the cell, according to the bright field image at the same location, was used to determine the whole cell signal ($\Phi_i$). Thus, the change of charge density on the whole cell membrane along with the cell electrical activity can be calculated by $\Delta\Phi = \Phi_i - \Phi_0$, where $\Phi_0$ is the whole cell signal at t = 0. Considering that the electrical activity of the cell is dominated by ion transport across the membrane, the summation of the activity of all ion channels on the membrane can be obtained in this mode.

For single channel detection (Fig. S3b), we introduce the concept of an accumulated temporal amplitude map (ATAP), wherein we sum all the STFT images into one image. The pixel value of ATAP is related to the change of surface charge density within observation time. Considering the blinking behavior of single ion channel and its continuously changing position due to membrane fluidity, active regions with larger pixel value in ATAP indicated a greater change of surface charge density, which may be caused by the activity of one or more ion channels. A region of interest (ROI) with 10 × 10 pixels was selected for each hot point that represents the ion channels active zone. The single channel signal, $\Delta A = A_i - A_0$, was determined by the amplitude of a 2D Gaussian function ($A_i$) fitting frame by frame to the spot at the corresponding ROI position in STFT image sequence (snaps shown in Fig. 3 and S3b), and the mean intensity within the ROI was used to determine the baseline ($A_0$). Dynamics related to channel opening and closing were determined based on the single channel signal from EM-iSCAT images.

**Data availability**

The data that support the results of this work are available from the corresponding authors upon reasonable request.

**Code availability**

All code used for data analysis is available from the corresponding authors upon reasonable request.

**Acknowledgements**

This work was mainly supported by the National Natural Science Foundation of China (22250009, 22034003, 21991080, 22261132510, 22174064), National Key R&D Program of China (Grant No. 2021YFA0910003), Excellent Research Program of Nanjing University (ZYJH004). We particularly thank PhD. Jiang Chen from Nanjing Drum Tower Hospital for providing technical guidance for cell transfection.

**Author contributions**

B. K., J.-J. X. and H.-Y. C. supervised the project. B. K. conceived the idea. B. K. and Q. L. designed the optical setup. Q. L. performed the experiments and analyzed the data. P. L. helped optimize the optical setup and contributed to data processing. B. K. and Q. L. wrote the manuscript. All authors discussed the results and commented on the manuscript.